\documentstyle[12pt]{article}
\def\1{{\chi}}

\begin{document}
%\centerline{Islamic University Journal\hfill Vol. 5, No. 2 June 1997}
%\hrule
%\vskip 2in
\title {{The average value inequality in sequential effect algebras}\thanks{This project is supported by Natural Science
Found of China (10771191 and 10471124).}}
%\vskip 1.5in
\author {Shen Jun$^{1,2}$, Wu Junde$^{1}$\date{}\thanks{E-mail: wjd@zju.edu.cn}}
\maketitle
$^1${\small\it Department of Mathematics, Zhejiang
University, Hangzhou 310027, P. R. China}

$^2${\small\it Department of Mathematics, Anhui Normal University,
Wuhu 241003, P. R. China}

\begin{abstract} {A sequential effect algebra $(E,0,1, \oplus , \circ)$ is an effect algebra on which a sequential product $\circ$ with certain physics properties is defined, in particular, sequential effect algebra is an important model for studying quantum measurement theory.
In 2005, Gudder asked the following problem: If $a, b\in (E,0,1,
\oplus , \circ)$ and $a\bot b$ and $a\circ b\bot a\circ b$, is it
the case that $2(a\circ b)\leq a^2\oplus b^2$ ? In this paper, we
construct an example to answer the problem negatively.}
\end{abstract}


{\bf MR(2000) Subject Classification.} {\small 81P15}

\vskip0.1in

{\it Effect algebra} was introduced in 1994 to model the quantum
logic which may be fuzzy or unsharp, to be precise, an effect
algebra is a system $(E,0,1, \oplus)$, where 0 and 1 are distinct
elements of $E$ and $\oplus$ is a partial binary operation on $E$
satisfying [1]:

(EA1) If $a\oplus b$ is defined, then $b\oplus a$ is defined and
$b\oplus a=a\oplus b$.

(EA2) If $a\oplus (b\oplus c)$ is defined, then $(a\oplus b)\oplus
c$ is defined and $$(a\oplus b)\oplus c=a\oplus (b\oplus c).$$

(EA3) For each $a\in E$, there exists a unique element $b\in E$ such
that $a\oplus b=1$.

(EA4) If $a\oplus 1$ is defined, then $a=0$.

\vskip 0.1 in

In an effect algebra $(E,0,1, \oplus)$, if $a\oplus b$ is defined,
we write $a\bot b$. If $a\bot a$, we denote $a\oplus a$ by $2a$. For
each $a\in (E,0,1, \oplus)$, it follows from (EA3) that there exists
a unique element $b\in E$ such that $a\oplus b=1$, we denote $b$ by
$a'$. If $a\wedge a'=0$, we say that $a$ is a {\it sharp element} of
$(E,0,1, \oplus)$ (see [2]). Let $a, b\in (E,0,1, \oplus)$, if there
exists a $c\in E$ such that $a\bot c$ and $a\oplus c=b$, then we say
that $a\leq b$. It follows from [1] that $\leq $ is a partial order
of $(E,0,1, \oplus)$ and satisfies that for each $a\in E$, $0\leq
a\leq 1$, $a\bot b$ iff $a\leq b'$.

\vskip 0.1 in

In 2001, in order to study quantum measurement theory, Gudder began
to consider the sequential product of two measurements $A$ and $B$
(see [3]). In 2002, Professors Gudder and Greechie introduced the
abstract sequential effect algebra structure, that is:

\vskip 0.1 in

A {\it sequential effect algebra} is an effect algebra $(E,0,1,
\oplus)$ and another binary operation $\circ $ defined on $(E,0,1,
\oplus)$ satisfying [4]:

(SEA1) The map $b\mapsto a\circ b$ is additive for each $a\in E$,
that is, if $b\bot c$, then $a\circ b\bot a\circ c$ and $a\circ
(b\oplus c)=a\circ b\oplus a\circ c$.

(SEA2) $1\circ a=a$ for each $a\in E$.

(SEA3) If $a\circ b=0$, then $a\circ b=b\circ a$.

(SEA4) If $a\circ b=b\circ a$, then $a\circ b'=b'\circ a$ and
$a\circ (b\circ c)=(a\circ b)\circ c$ for each $c\in E$.

(SEA5) If $c\circ a=a\circ c$ and $c\circ b=b\circ c$, then
$c\circ(a\circ b)=(a\circ b)\circ c$ and $c\circ(a\oplus b)=(a\oplus
b)\circ c$ whenever $a\bot b$.

\vskip 0.1 in

Let $(E,0,1, \oplus, \circ)$ be a sequential effect algebra. Then
the operation $\circ$ is said to be a sequential product on $(E,0,1,
\oplus, \circ)$. If $a, b\in (E,0,1, \oplus, \circ)$ and $a\circ
b=b\circ a$, then we say that $a$ and $b$ is {\it sequentially
independent} and denoted by $a|b$ (see [4]). If $a\in (E,0,1,
\oplus, \circ)$, we denote $a\circ a$ by $a^2$, it follows from ([4,
Lemma 3.2]) that $a$ is a sharp element of $(E,0,1, \oplus , \circ)$
iff $a^2=a$. We denote the set of all sharp elements in $(E,0,1,
\oplus, \circ)$ by $E_s$.

\vskip 0.1 in

In 2005, in order to motivate the study of sequential effect algebra
theory, Professor Gudder presented 25 important and interesting
problems, the 23th problem asked ([5]): If $a, b\in (E,0,1, \oplus ,
\circ)$ and $a\bot b$ and $a\circ b\bot a\circ b$, is it the case
that $2(a\circ b)\leq a^2\oplus b^2$ ? In this paper, we construct
an example to answer the problem negatively.

\vskip 0.1 in

At first, we show that the above average value inequality does hold
in the underlying sequential effect algebras under some additional
conditions. That is:

\vskip 0.1 in

{\bf Proposition 1.} If $(E,0,1, \oplus , \circ)$ is a sequential
effect algebra, $a,b\in E$, $a^2\perp b^2$(a sufficient condition
for this is $a\perp b$), $a\leq b$(or $b\leq a$) and $a|b$, then
$(a\circ b)\perp(a\circ b)$ and $2(a\circ b)\leq a^2\oplus b^2$.

{\bf Proof.} Since $a\leq b$, there exists a $c\in E$ such that
$a\oplus c=b$. Since $a|b$, it follows that $c|b$ (see [4] Lemma
3.1(v)).

$c\circ b=c\circ(a\oplus c)=(c\circ a)\oplus c^2$.

$b^2=b\circ(a\oplus c)=(b\circ a)\oplus (b\circ c)=(a\circ b)\oplus
(c\circ b)=(a\circ b)\oplus (c\circ a)\oplus c^2$.

Since $a^2\perp b^2$, $a^2\oplus b^2=a^2\oplus(a\circ b)\oplus
(c\circ a)\oplus c^2$.

While $a\circ b=a\circ(a\oplus c)=a^2\oplus (a\circ c)=a^2\oplus
(c\circ a)$, so $a^2\oplus b^2=(a\circ b)\oplus(a\circ b)\oplus
c^2$.

It follows that $(a\circ b)\perp(a\circ b)$ and $2(a\circ b)\leq
a^2\oplus b^2$.

Finally, if $a\perp b$, it follows from $a^2\leq a$ and $b^2\leq b$
that $a^2\perp b^2$. The proposition is proved.

\vskip 0.1 in

{\bf Proposition 2.} If $(E,0,1, \oplus , \circ)$ is a sequential
effect algebra, $a,b\in E$, $a\perp b$, $a\in E_s$(or $b\in E_s$),
then $(a\circ b)\perp(a\circ b)$ and $2(a\circ b)\leq a^2\oplus
b^2$.

{\bf Proof.} Since $a\perp b$ and $a\in E_s$, it follows that
$a\circ b=0$ (see [4] Lemma 3.3(ii)), so $2(a\circ b)\leq a^2\oplus
b^2$.

\vskip 0.1 in

Now, we construct a sequential effect algebra to show that the above
average value inequality does not always hold.

\vskip 0.1 in

In this paper, we denote ${\mathbf{Z}}$ the integer set,
${\mathbf{N}}$ the nonnegative integer set and ${\mathbf{N}}^+$
the positive integer set.

Let $E_0=\{0,1,a_n,b_n,c_{i,k,m},d_{i,k,m}|\ n\in
{\mathbf{N}}^+,i,k\in {\mathbf{N}}\ and\ i^2+k^2\neq 0,m\in
{\mathbf{Z}}\}$. For simplicity, in the sequel, unless specified,
the subindex of respective elements will always take values in the
corresponding default sets. To be accurately, when we write
$a_n,b_n$, $n$ always take values in ${\mathbf{N}}^+$, when we write
$c_{i,k,m},d_{i,k,m}$, $i,k$ always take values in ${\mathbf{N}}$
and $i^2+k^2\neq 0$ and $m$ always take values in ${\mathbf{Z}}$.

We define a partial binary operation $\oplus$ on $E_0$ as
follows(when we write $x\oplus y=z$, we always mean $x\oplus
y=z=y\oplus x$):

For each $x\in E_0$, $0\oplus x=x$,

$a_n\oplus a_m=a_{n+m}$, $a_n\oplus c_{i,k,m}=c_{i,k,n+m}$,
$a_n\oplus d_{i,k,m}=d_{i,k,m-n}$, $c_{i,k,m}\oplus
c_{r,s,t}=c_{i+r,k+s,m+t}$.

For $n<m$, $a_n\oplus b_m=b_{m-n}$, $a_n\oplus b_n=1$.

For $i\leq r\ and\ k\leq s\ and\ (r-i)^2+(s-k)^2\neq 0$.
$c_{i,k,m}\oplus d_{r,s,t}=d_{r-i,s-k,t-m}$.

For $i=r\ and\ k=s\ and\ m<t$, $c_{i,k,m}\oplus d_{r,s,t}=b_{t-m}$.

For $i=r\ and\ k=s\ and\ m=t$, $c_{i,k,m}\oplus d_{r,s,t}=1$.

No other $\oplus$ operation is defined.

Next, we define a binary operation  $\circ$  on $E_0$ as
follows(when we write $x\circ y=z$, we always mean $x\circ
y=z=y\circ x$):

For each $x\in E_0$, $0\circ x=0$, $1\circ x=x$,

$a_n\circ a_m=0$, $a_n\circ b_m=a_n$, $b_n\circ b_m=b_{m+n}$,
$a_n\circ c_{i,k,m}=0$, $c_{i,k,m}\circ b_n=c_{i,k,m}$, $a_n\circ
d_{i,k,m}=a_n$, $b_n\circ d_{i,k,m}=d_{i,k,m+n}$, $d_{i,k,m}\circ
d_{r,s,t}=d_{i+r,k+s,m+t-is-kr}$, $c_{i,k,m}\circ
d_{r,s,t}=c_{i,k,m-is-kr}$, $c_{i,k,m}\circ
c_{r,s,t}=a_{is+kr}(when\ is+kr\neq 0)\ or\ 0(when\ is+kr=0)$.

{\bf Proposition 3.} $(E_0,0,1, \oplus , \circ)$ is a sequential
effect algebra.

{\bf Proof.} First we verify that $(E_0,0,1, \oplus)$ is an effect
algebra.

(EA1) and (EA4) are trivial.

We verify (EA2), we omit the trivial cases about 0,1:

$a_n\oplus (a_m\oplus a_{k})=(a_n\oplus a_m)\oplus a_{k}=a_{k+m+n}$.

$a_n\oplus (a_m\oplus c_{i,j,k})=(a_n\oplus a_m)\oplus
c_{i,j,k}=c_{i,j,k+m+n}$.

$a_n\oplus (a_m\oplus d_{i,j,k})=(a_n\oplus a_m)\oplus
d_{i,j,k}=d_{i,j,k-m-n}$.

$a_n\oplus (c_{r,s,t}\oplus c_{i,j,k})=(a_n\oplus c_{r,s,t})\oplus
c_{i,j,k}=c_{i+r,s+j,k+t+n}$.

$c_{l,m,n}\oplus (c_{r,s,t}\oplus c_{i,j,k})=(c_{l,m,n}\oplus
c_{i,j,k})\oplus c_{r,s,t}=c_{i+l+r,j+m+s,k+n+t}$.

Each $a_n\oplus (a_m\oplus b_{k})$ or $(a_n\oplus a_m)\oplus b_{k}$
is defined iff $n+m\leq k$, at this case, $a_n\oplus (a_m\oplus
b_{k})=(a_n\oplus a_m)\oplus b_{k}=b_{k-m-n}(when\ m+n<k)\ or\
1(when\ m+n=k)$.

Each $a_n\oplus (c_{r,s,t}\oplus d_{i,j,k})$ or $(a_n\oplus
c_{r,s,t})\oplus d_{i,j,k}$ or $(a_n\oplus d_{i,j,k})\oplus
c_{r,s,t}$ is defined iff one of the following two conditions is
satisfied:

(1) $r\leq i\ and\ s\leq j\ and\ (i-r)^2+(j-s)^2\neq 0$, at this
case, $a_n\oplus (c_{r,s,t}\oplus d_{i,j,k})=(a_n\oplus
c_{r,s,t})\oplus d_{i,j,k}=(a_n\oplus d_{i,j,k})\oplus
c_{r,s,t}=d_{i-r,j-s,k-t-n}$;

(2) $r=i\ and\ s=j\ and\ n+t\leq k$, at this case, $a_n\oplus
(c_{r,s,t}\oplus d_{i,j,k})=(a_n\oplus c_{r,s,t})\oplus
d_{i,j,k}=(a_n\oplus d_{i,j,k})\oplus c_{r,s,t}=b_{k-t-n}(when\
n+t<k)\ or\ 1(when\ n+t=k)$.

Each $c_{l,m,n}\oplus (c_{r,s,t}\oplus d_{i,j,k})$ or
$(c_{l,m,n}\oplus c_{r,s,t})\oplus d_{i,j,k}$ is defined iff one of
the following two conditions is satisfied:

(1) $l+r\leq i\ and\ m+s\leq j\ and\ (i-l-r)^2+(j-m-s)^2\neq 0$, at
this case, $c_{l,m,n}\oplus (c_{r,s,t}\oplus
d_{i,j,k})=(c_{l,m,n}\oplus c_{r,s,t})\oplus
d_{i,j,k}=d_{i-l-r,j-m-s,k-t-n}$;

(2) $l+r=i\ and\ m+s=j\ and\ n+t\leq k$, at this case,
$c_{l,m,n}\oplus (c_{r,s,t}\oplus d_{i,j,k})=(c_{l,m,n}\oplus
c_{r,s,t})\oplus d_{i,j,k}=b_{k-t-n}(when\ n+t<k)\ or\ 1(when\
n+t=k)$.

We verify (EA3): $a_n\oplus b_n=1$, $c_{i,k,m}\oplus d_{i,k,m}=1$.

So $(E_0,0,1, \oplus)$ is an effect algebra.

We now verify that $(E_0,0,1, \oplus , \circ)$ is a sequential
effect algebra.

(SEA2) and (SEA3) and (SEA5) are trivial.

We verify (SEA1), we omit the trivial cases about 0,1:

$a_n\circ (a_m\oplus a_{k})=a_n\circ a_m\oplus a_n\circ a_{k}=0$,

$b_n\circ (a_m\oplus a_{k})=b_n\circ a_m\oplus b_n\circ
a_{k}=a_{m+k}$,

$c_{r,s,t}\circ (a_m\oplus a_{k})=c_{r,s,t}\circ a_m\oplus
c_{r,s,t}\circ a_{k}=0$,

$d_{r,s,t}\circ (a_m\oplus a_{k})=d_{r,s,t}\circ a_m\oplus
d_{r,s,t}\circ a_{k}=a_{m+k}$.

$a_n\circ (a_m\oplus c_{r,s,t})=a_n\circ a_m\oplus a_n\circ
c_{r,s,t}=0$,

$b_n\circ (a_m\oplus c_{r,s,t})=b_n\circ a_m\oplus b_n\circ
c_{r,s,t}=c_{r,s,m+t}$,

$c_{x,y,z}\circ (a_m\oplus c_{r,s,t})=c_{x,y,z}\circ a_m\oplus
c_{x,y,z}\circ c_{r,s,t}=a_{xs+yr}(when\ xs+yr\neq\ 0)\ or\ 0(when\
xs+yr=0)$,

$d_{x,y,z}\circ (a_m\oplus c_{r,s,t})=d_{x,y,z}\circ a_m\oplus
d_{x,y,z}\circ c_{r,s,t}=c_{r,s,m+t-xs-yr}$.

$a_n\circ (a_m\oplus d_{r,s,t})=a_n\circ a_m\oplus a_n\circ
d_{r,s,t}=a_n$,

$b_n\circ (a_m\oplus d_{r,s,t})=b_n\circ a_m\oplus b_n\circ
d_{r,s,t}=d_{r,s,n+t-m}$,

$c_{x,y,z}\circ (a_m\oplus d_{r,s,t})=c_{x,y,z}\circ a_m\oplus
c_{x,y,z}\circ d_{r,s,t}=c_{x,y,z-xs-yr}$,

$d_{x,y,z}\circ (a_m\oplus d_{r,s,t})=d_{x,y,z}\circ a_m\oplus
d_{x,y,z}\circ d_{r,s,t}=d_{x+r,y+s,z+t-m-xs-yr}$.

$a_n\circ (c_{x,y,z}\oplus c_{r,s,t})=a_n\circ c_{x,y,z}\oplus
a_n\circ c_{r,s,t}=0$,

$b_n\circ (c_{x,y,z}\oplus c_{r,s,t})=b_n\circ c_{x,y,z}\oplus
b_n\circ c_{r,s,t}=c_{x+r,y+s,z+t}$,

$c_{i,k,m}\circ (c_{x,y,z}\oplus c_{r,s,t})=c_{i,k,m}\circ
c_{x,y,z}\oplus c_{i,k,m}\circ c_{r,s,t}=a_{i(y+s)+k(x+r)}(when\
i(y+s)+k(x+r)\neq 0)\ or\ 0(when\ i(y+s)+k(x+r)=0)$,

$d_{i,k,m}\circ (c_{x,y,z}\oplus c_{r,s,t})=d_{i,k,m}\circ
c_{x,y,z}\oplus d_{i,k,m}\circ
c_{r,s,t}=c_{x+r,y+s,z+t-i(y+s)-k(x+r)}$.

For $m\leq k$,

$a_n\circ (a_m\oplus b_{k})=a_n\circ a_m\oplus a_n\circ b_{k}=a_n$,

$b_n\circ (a_m\oplus b_{k})=b_n\circ a_m\oplus b_n\circ
b_{k}=b_{n+k-m}$,

$c_{x,y,z}\circ (a_m\oplus b_{k})=c_{x,y,z}\circ a_m\oplus
c_{x,y,z}\circ b_{k}=c_{x,y,z}$,

$d_{x,y,z}\circ (a_m\oplus b_{k})=d_{x,y,z}\circ a_m\oplus
d_{x,y,z}\circ b_{k}=d_{x,y,z+k-m}$.

For $i\leq r\ and\ k\leq s\ and\ (r-i)^2+(s-k)^2\neq 0$,

$a_n\circ (c_{i,k,m}\oplus d_{r,s,t})=a_n\circ c_{i,k,m}\oplus
a_n\circ d_{r,s,t}=a_n$,

$b_n\circ (c_{i,k,m}\oplus d_{r,s,t})=b_n\circ c_{i,k,m}\oplus
b_n\circ d_{r,s,t}=d_{r-i,s-k,n+t-m}$,

$c_{x,y,z}\circ (c_{i,k,m}\oplus d_{r,s,t})=c_{x,y,z}\circ
c_{i,k,m}\oplus c_{x,y,z}\circ d_{r,s,t}=c_{x,y,z-x(s-k)-y(r-i)}$,

$d_{x,y,z}\circ (c_{i,k,m}\oplus d_{r,s,t})=d_{x,y,z}\circ
c_{i,k,m}\oplus d_{x,y,z}\circ
d_{r,s,t}=d_{x+r-i,y+s-k,z+t-m-x(s-k)-y(r-i)}$.

For $i=r\ and\ k=s\ and\ m\leq t$,

$a_n\circ (c_{i,k,m}\oplus d_{r,s,t})=a_n\circ c_{i,k,m}\oplus
a_n\circ d_{r,s,t}=a_n$,

$b_n\circ (c_{i,k,m}\oplus d_{r,s,t})=b_n\circ c_{i,k,m}\oplus
b_n\circ d_{r,s,t}=b_{n+t-m}$,

$c_{x,y,z}\circ (c_{i,k,m}\oplus d_{r,s,t})=c_{x,y,z}\circ
c_{i,k,m}\oplus c_{x,y,z}\circ d_{r,s,t}=c_{x,y,z}$,

$d_{x,y,z}\circ (c_{i,k,m}\oplus d_{r,s,t})=d_{x,y,z}\circ
c_{i,k,m}\oplus d_{x,y,z}\circ d_{r,s,t}=d_{x,y,z+t-m}$.

We verify (SEA4), we omit the trivial cases about 0,1:

$a_n\circ (a_m\circ a_{k})=(a_n\circ a_m)\circ a_{k}=0$,

$a_n\circ (a_m\circ b_{k})=b_{k}\circ (a_n\circ a_m)=a_m\circ
(a_n\circ b_{k})=0$,

$a_n\circ (a_m\circ c_{r,s,t})=c_{r,s,t}\circ (a_n\circ
a_m)=a_m\circ (a_n\circ c_{r,s,t})=0$,

$a_n\circ (a_m\circ d_{r,s,t})=d_{r,s,t}\circ (a_n\circ
a_m)=a_m\circ (a_n\circ d_{r,s,t})=0$,

$a_n\circ (b_m\circ b_{k})=b_{k}\circ (a_n\circ b_m)=b_m\circ
(a_n\circ b_{k})=a_n$,

$a_n\circ (b_m\circ c_{r,s,t})=c_{r,s,t}\circ (a_n\circ
b_m)=b_m\circ (a_n\circ c_{r,s,t})=0$,

$a_n\circ (b_m\circ d_{r,s,t})=d_{r,s,t}\circ (a_n\circ
b_m)=b_m\circ (a_n\circ d_{r,s,t})=a_n$,

$a_n\circ (c_{i,k,m}\circ c_{r,s,t})=c_{r,s,t}\circ (a_n\circ
c_{i,k,m})=c_{i,k,m}\circ (a_n\circ c_{r,s,t})=0$,

$a_n\circ (c_{i,k,m}\circ d_{r,s,t})=d_{r,s,t}\circ (a_n\circ
c_{i,k,m})=c_{i,k,m}\circ (a_n\circ d_{r,s,t})=0$,

$a_n\circ (d_{i,k,m}\circ d_{r,s,t})=d_{r,s,t}\circ (a_n\circ
d_{i,k,m})=d_{i,k,m}\circ (a_n\circ d_{r,s,t})=a_n$,

$b_n\circ (b_m\circ b_{k})=b_{k}\circ (b_n\circ b_m)=b_{m+n+k}$,

$b_n\circ (b_m\circ c_{r,s,t})=c_{r,s,t}\circ (b_n\circ
b_m)=b_m\circ (b_n\circ c_{r,s,t})=c_{r,s,t}$,

$b_n\circ (b_m\circ d_{r,s,t})=d_{r,s,t}\circ (b_n\circ
b_m)=b_m\circ (b_n\circ d_{r,s,t})=d_{r,s,n+m+t}$,

$b_n\circ (c_{i,k,m}\circ c_{r,s,t})=c_{r,s,t}\circ (b_n\circ
c_{i,k,m})=c_{i,k,m}\circ (b_n\circ c_{r,s,t})=a_{is+kr}(when\
is+kr\neq 0)\ or\ 0(when\ is+kr=0)$,

$b_n\circ (c_{i,k,m}\circ d_{r,s,t})=d_{r,s,t}\circ (b_n\circ
c_{i,k,m})=c_{i,k,m}\circ (b_n\circ d_{r,s,t})=c_{i,k,m-is-kr}$,

$b_n\circ (d_{i,k,m}\circ d_{r,s,t})=d_{r,s,t}\circ (b_n\circ
d_{i,k,m})=d_{i,k,m}\circ (b_n\circ
d_{r,s,t})=d_{i+r,k+s,n+m-t-is-kr}$,

$c_{x,y,z}\circ (c_{i,k,m}\circ c_{r,s,t})=c_{r,s,t}\circ
(c_{x,y,z}\circ c_{i,k,m})=0$,

$c_{x,y,z}\circ (c_{i,k,m}\circ d_{r,s,t})=d_{r,s,t}\circ
(c_{x,y,z}\circ c_{i,k,m})=c_{i,k,m}\circ (c_{x,y,z}\circ
d_{r,s,t})=a_{xk+yi}(when\ xk+yi\neq 0)\ or\ 0(when\ xk+yi=0)$,

$c_{x,y,z}\circ (d_{i,k,m}\circ d_{r,s,t})=d_{r,s,t}\circ
(c_{x,y,z}\circ d_{i,k,m})=d_{i,k,m}\circ (c_{x,y,z}\circ
d_{r,s,t})=c_{x,y,z-x(k+s)-y(i+r)}$,

$d_{x,y,z}\circ (d_{i,k,m}\circ d_{r,s,t})=d_{r,s,t}\circ
(d_{x,y,z}\circ
d_{i,k,m})$$=d_{x+i+r,y+k+s,z+m+t-(is+kr+xk+xs+yi+yr)}$.

So $(E_0,0,1, \oplus , \circ)$ is a sequential effect algebra.

Our main result is:

{\bf Theorem 1.} The average value inequality does not always hold
in sequential effect algebras.

{\bf Proof.} In fact, in $(E_0,0,1, \oplus , \circ)$,
$c_{1,0,0}\perp c_{0,1,0}$, $c_{1,0,0}\oplus c_{0,1,0}=c_{1,1,0}$.
$c_{1,0,0}\circ c_{0,1,0}=a_{1}$, $a_{1}\perp a_{1}$, $a_{1}\oplus
a_{1}=a_{2}$. But $2(c_{1,0,0}\circ c_{0,1,0})=2a_{1}=a_{1}\oplus
a_{1}=a_{2}$, $(c_{1,0,0})^2=c_{1,0,0}\circ c_{1,0,0}=0$,
$(c_{0,1,0})^2=c_{0,1,0}\circ c_{0,1,0}=0$, so $2(c_{1,0,0}\circ
c_{0,1,0})\not\leq (c_{1,0,0})^2\oplus (c_{0,1,0})^2$.

{\bf Remarks.} Recently, the 2th problem, the 3th problem, the 17th
problem and the 20th problem of Gudder have also been answered
([6-9]).

\vskip0.2in

\centerline{\bf References}

\vskip0.2in

\noindent [1]. Foulis, D J, Bennett, M K. Effect algebras and
unsharp quantum logics. Found Phys 24 (1994), 1331-1352.

\noindent [2]. Gudder, S. Sharply dominating effect algebras. Tatra
Mt. Math. Publ., 15(1998), 23-30.

\noindent [3]. Gudder, S, Nagy, G. Sequential quantum measurements.
J. Math. Phys. 42(2001), 5212-5222.

\noindent [4]. Gudder, S, Greechie, R. Sequential products on effect
algebras. Rep. Math. Phys.  49(2002), 87-111.

\noindent [5]. Gudder, S. Open problems for sequential effect
algebras. Inter. J. Theory. Phys. 44 (2005), 2219-2230.

\noindent [6]. Weihua Liu, Junde Wu. The Uniqueness Problem of
Sequence Product on Operator Effect Algebra ${\cal E}(H)$. J. Physi.
A (Accepted to appear).

\noindent [7]. Jun Shen, Junde Wu. Not each sequential effect
algebra is sharply dominating. Physics Letter A (Accepted to
appear).

\noindent [8]. Jun Shen, Junde Wu. Remarks on the sequential effect
algebras. Report Math. Physi. (Accepted to appear).

\noindent [9]. Jun Shen, Junde Wu. The square root is not unique in
sequential effect algebras. (To appear).

\end{document}